\documentstyle[preprint,aps]{revtex}
\begin{document}
\def\baselinestretch{1.5}

\title{Low temperature resistivity in a 
nearly half-metallic ferromagnet}
\author{Xindong Wang$^1$ and X.-G. Zhang$^2$\\
{\small $^1$ Metals and Ceramics Division, Oak Ridge National Laboratory,
Oak Ridge, TN 37831-6114} \\
{\small $^2$ Computational Physics and Engineering Division, 
Oak Ridge National Laboratory,\\ Oak Ridge, TN 37831}}
\maketitle

\abstract{We consider electron transport in a nearly half-metallic ferromagnet, 
in which the minority
spin electrons close to the band edge at the Fermi energy are
Anderson-localized due to disorder.
For the case of spin-flip scattering of the conduction electrons
due to the absorption and emission of
magnons, the Boltzmann equation is exactly soluble to the linear order.
From this solution we calculate the
temperature dependence of the resistivity due to single magnon processes 
at sufficiently low temperature, namely $k_BT\ll D/L^2$,
where $L$ is the Anderson localization length and $D$ is the magnon stiffness.
And
depending on the details of the minority spin density of states at 
the Fermi level, 
we find a $T^{1.5}$ or $T^{2}$ scaling behavior for resistivity.
Relevance to the 
doped perovskite manganite systems is discussed.
 }
\newline
\today
\maketitle
\newpage
The doped manganite perovskites, R$_{1-x}$A$_{x}$MnO$_3$, where
R is a rare earth element, and A is some divalent ions,
exhibit very interesting transport properties. In addition to the
well-known phenomenon of ``colossal'' magnetoresistance at the doping
range of around $x\approx 1/3$, it has also been observed that below Curie temperature
the electric resistivities can be fitted with
a $T^{2.5}$ dependence at the same doping range\cite{schiffer}.
A good theoretical description of electron transport for magnetic systems should 
certainly be able to account for the contributions from the electron-magnon scattering.
In conventional metallic ferromagnet, it has long
been established\cite{japanese,Mannari} that the resistivity due
to absorption and emission of a single magnon gives rise to a
$T^2$ temperature dependence. 
The manganite perovskites, however, 
have been proposed\cite{pickett} to be almost half-metallic
ferromagnets, where the Fermi energy lies near the bottom of the minority spin 
bands.
For a truly half-metallic system,  Kubo and Ohata\cite{kubo} have shown
that while single magnon processes are exponentially suppressed by
a factor $\exp(-E_g/k_BT)$ where $E_g$ is the minority spin band gap at the
Fermi energy, the double
magnon processes can lead to a $T^{4.5}$ temperature dependence. 
The ramification of the fact that this gap may actually be
zero and the Fermi energy passes through a region of low 
density of states in the minority spin channel\cite{pickett}
in these manganites, however,
is still left unexplored.

In this Letter, we propose a model for the electron transport in this type of 
nearly-half metallic ferromagnet
(NHMF).  In this model, there is no true band gap at the Fermi energy for the
minority spin electrons. Instead, the minority spin electrons do not conduct
current due to Anderson-localization driven by disorder. Therefore, the system
is half-metallic as far as transport is concerned. But there is a finite
electronic density of states at the Fermi energy for the minority spin. This
allows spin-flip scatterings involving only single magnons to occur.

The relevant magnon contributions at low temperatures
are from the long wavelength magnons, for which the energy dispersion can
be accurately described by $\omega({\bf q})=Dq^2$,
with $D$ being the stiffness.
The squared matrix element for 
the absorption or emission of a long wavelength magnon with wavevector ${\bf q}$ 
is given by,
\begin{eqnarray}
M_l({\bf k},{\bf q})= \frac{2J^2 S}{N^2}|\phi_l({\bf k}-{\bf q})|^2
\approx \frac{2J^2 S}{N^2}\left(\frac{L}{a}\right)^{3}\exp(-L^2|{\bf k}-{\bf q}|^2),
\label{anderson}
\end{eqnarray}
where $N$ is total number of lattice sites,
$\phi_l({\bf k})$ are the Fourier components of the 
localized minority spin electron wavefunctions, $\phi_l$, the factor $2J^2 S$
comes from ferromagnetic s-d 
type interaction, $a^3$ is the volume of crystal unit cell,
and $L$ is the Anderson localization length. 
In the limit of $L\to\infty$, there is no Anderson localization, and
$M({\bf k},{\bf q})$ 
simply gives a delta function corresponding to the momentum conservation.
On the other hand, if $L$ is sufficiently small, or alternatively the
temperature is sufficiently low, so that
$k_BT\ll D/L^2$, then the ${\bf k}$-dependence of
$|\phi_l({\bf k})|^2$ is weak, and the matrix elements $M_l({\bf k},{\bf q})$ can
be approximated by a constant 
\begin{eqnarray}
M_l=\frac{\mu_l}{N^2}
\approx \frac{2J^2 S}{N^2}\left(\frac{L}{a}\right)^{3}\exp(-L^2 k_F^2)
\label{M_l}
\end{eqnarray}
for each localized state. 

We wish to emphasize the importance of Anderson localization
to our NHMF model.  The
itinerant model obtained from our model by taking the $L\to\infty$ limit
is {\it very} different from the model used in
Refs.\cite{japanese,Mannari} where no
conduction electron exchange splitting is considered. In our case, without
Anderson localization,
an additional delta function for
the momentum conservation will force the magnon energy to be at least
$E_{\rm stoner}$, the Stoner threshold. This
suppresses the single magnon process by
a factor of $\exp(-E_{\rm stoner}/k_BT)$. 
Without Anderson localization,
the phase space
that satisfies both the energy and the momentum conservation is
exponentially small ($E_{\rm stoner}>0$)
even when the minority spin band
gap at Fermi energy is zero.
Therefore, Anderson 
localization of minority spin electrons is necessary for the contribution
of the single magnon process to be significant to the low temperature
resistivity. We show later that the Anderson localization also allows the 
exact solution of the model to linear order.

Let us now turn to the Boltzmann equation for majority spin electrons,
\begin{eqnarray}
-e{\bf v}_1({\bf k})F'_1[E_1({\bf k})]\cdot{\cal E}=
\left(\frac{ d F_1[E_1({\bf k})]} {dt}\right)_{\rm col},
\label{boltz}
\end{eqnarray}
where $e=|e|$, $\hbar{\bf v}({\bf k})=\nabla_{\bf k} E_1({\bf k})$, ${\cal E}$ is the applied
external field, and $F_1[E_1({\bf k})]$ is the distribution function of the
majority spin electrons, and the subscript $1$ denotes the majority spin.
To calculate the collision term
for the single magnon spin-flip processes,
we use the low temperature limit of $M_l({\bf k},{\bf q})$, Eq. (\ref{M_l}).
The collision term is thus
\begin{eqnarray}
\left(\frac{ d F_1[E_1({\bf k})]}{dt}\right)_{\rm col}=\frac{2\pi}{\hbar}
\sum_l\sum_{\bf q} \{
F_2[E_1({\bf k})+\omega({\bf q})](1-F_1[E_1({\bf k})])(n[\omega({\bf q})]+1)\nonumber\\
-F_1[E_1({\bf k})](1-F_2[E_1({\bf k})+\omega({\bf q})]) n[\omega({\bf q})]\}
M_l\delta[E_l-E_1({\bf k})-\omega({\bf q})],
\label{col}
\end{eqnarray}
where the first term is due to 
the emission of a magnon by a minority spin electron and the second term is due
to the absorption of a magnon by a majority spin electron, and the subscript
$2$ denotes the minority spin.
Here $n(x)=1/[\exp(x/k_BT)-1]$ is the Bose-Einstein distribution function,
$F_2(x)$ is the 
distribution function for the minority spin electrons, and $E_l$ is the
energy of the localized minority
state $\phi_l$.

Because $\phi_l$'s are localized, the distribution function for the minority
spin electrons are unchanged by the external field to the linear order in
${\cal E}$\cite{notes}, i.e.,
$F_2(x)=f(x)$, where $f(x)=1/\{\exp[(x-E_F)/k_BT]+1\}$ is the
equilibrium Fermi distribution function, and $E_F$ is the
Fermi energy. As is shown below, the fact that the minority spin distribution 
is unchanged to linear order allows an exact linear order
solution of the Boltzmann equation

We define
\begin{eqnarray}
-e\Delta({\bf k})=\left\{F_1[E_1({\bf k})]-f[E_1({\bf k})]\right\},
\end{eqnarray}
so that the current is given by
\begin{eqnarray}
{\bf j}=\frac{e^2}{8\pi^3} \int d{\bf k} {\bf v}_1({\bf k})\Delta({\bf k}).
\label{current}
\end{eqnarray}
Thus 
Eq. (\ref{col}) gives
\begin{eqnarray}
\left(\frac{ d F_1[E_1({\bf k})]}{dt}\right)_{\rm col} &=&
\frac{2\pi}{\hbar}\sum_l\sum_{\bf q}
e\Delta({\bf k}) M_l
\frac{f[E_1({\bf k})+\omega({\bf q})]}{f[E_1({\bf k})]
}\nonumber\\
&&\exp\left[\frac{\omega({\bf q})}{k_BT}\right]n[\omega({\bf q})]\delta[E_l-
E_1({\bf l})-\omega({\bf q})],
\label{coll2}
\end{eqnarray}
Combining Eqs. (\ref{boltz}) and (\ref{coll2}), and keeping the terms
linear to the external field, we get
\begin{eqnarray}
-{\bf v}_1({\bf k}) f'[E_1({\bf k})]\cdot{\cal E}&=& 
\frac{2\pi}{\hbar}\sum_l\sum_{\bf q}
M_l \Delta({\bf k})
\frac{f[E_1({\bf k})+\omega({\bf q})]}{
f[E_1({\bf k})]}\nonumber\\
&&\exp\left[\frac{\omega({\bf q})}{k_BT}\right]n[\omega({\bf q})]\delta[E_l-
E_1({\bf k})-\omega({\bf q})].
\end{eqnarray}
Upon solving for $\Delta({\bf k})$, we find,
\begin{eqnarray}
\Delta({\bf k})&=&
-\frac{\hbar}{2\pi}{\bf v}_1({\bf k}) f'[E_1({\bf k})]\cdot{\cal E}
\nonumber\\
&&
\left\{\frac{\bar{\mu}}{N}\sum_{\bf q} \frac{f[E_1({\bf k})+\omega({\bf q})]}
{f[E_1({\bf k})]}
\exp\left[\frac{\omega({\bf q})}{k_BT}\right]n[\omega({\bf q})]
g_2[E_1({\bf k})+\omega({\bf q})]\right\}^{-1},
\end{eqnarray}
where we have assumed that $M_l$ are not very different for different
localized states, with $\bar{\mu}$ being the average of the matrix elements
$\mu_l$, and
$g_2(E)=\frac{1}{N}\sum_l \delta(E_l-E)$ for the minority density of states.
We define variables $x=[E_1({\bf k})-E_F]/k_BT$ and $y=\omega({\bf q})/k_BT$,
to simplify the above equation into
\begin{eqnarray}
\Delta({\bf k})=
\frac{\hbar}{2\pi}{\bf v}_1({\bf k}) \frac{1}{k_BT}\frac{\exp(x)}{[\exp(x)+1]^3}
\cdot{\cal E}
\left\{\frac{\bar{\mu}}{4\pi^2}\left(\frac{k_BT}{D/a^2}\right)^{1.5}
I(x,k_BT)
\right\}^{-1}.
\end{eqnarray}
where
\begin{eqnarray}
I(x,k_BT)=\int_0^\infty dy \sqrt{y}\frac{g_2[E_F+k_BT(x+y)]}
{[\exp(x+y)+1][1-\exp(-y)]}.
\end{eqnarray}
Multiplying the above equation by ${\bf v}({\bf k})$ and integrating over ${\bf k}$, 
and also noting that the integrand is sharply peaked at $E_1({\bf k})=E_F$,
we get (cf Eq.\ref{current})
\begin{eqnarray}
\rho=\frac{3\bar{\mu}a^3}{2\pi\hbar e^2 g_1(E_F)v^2_1(E_F)}
\left(\frac{k_BT}{D/a^2}\right)^{1.5}
\left\{\int_{-\infty}^\infty dx \frac{\exp(x)}{[\exp(x)+1]^3I(x,k_BT)}
\right\}^{-1},
\end{eqnarray}
where the majority spin density of states $g_1(E)$ is defined as,
\begin{eqnarray}
g_1(E)=\frac{a^3}{8\pi^3}\int d{\bf k} \delta [E-E_1({\bf k})].
\end{eqnarray}
We assume that the minority density of states, $g_2$, scales near the
Fermi energy as,
$g_2(E_F+E)\sim E^{\alpha}$.
This leads to the scaling behavior of the resistivity,
$\rho\sim T^{1.5+\alpha}$.

It seems that the sensible possibilities for $\alpha$ in our NHMF model
is either zero or 
$0.5$. The former corresponds to a finite density of states for minority
spin electrons (even though they are localized) and the latter
to the situation that the Fermi surface just touches the edge of the
minority spin bands, and it is assumed that Anderson localization does not
change the density of states profile. Thus we find that in the low temperature
limit, i.e., $k_BT\ll D/L^2$, the resistivity due to single magnon scattering
scales with the temperature with an exponent between $1.5$ and $2$.
The $T^{2.5}$ temperature dependence reported experimentally\cite{schiffer}
seems to correspond to a higher temperature range, and thus may not be
dominated by the single magnon processes. However, since the temperature
dependence of the contribution from the single magnon processes has a lower
exponent than electron-electron scattering ($T^2$) or two magnon processes
($T^{4.5}$), single magnon scattering may become important at a sufficiently
low temperature.

The disorder in the NHMF materials such as doped perovskite manganites is due to either
Mn-O-Mn bond bending or static spin canting\cite{de genne}.
When
an external magnetic field is applied, the disorder is expected to be reduced.
Consequently, the Anderson localization length will increase with the
applied magnetic field. This in turn reduces the matrix element
$M$ for the single magnon scattering (cf Eq.\ref{anderson}), and reduces the
resistivity of the material.
Thus even at low temperature, these NHMF materials should also
have a large magnetoresistance.

XDW thank Dr. Jiandi Zhang for bringing my attention to Ref\cite{schiffer}.
This research is sponsored by the U.S. Department of Energy 
under Contract DE-AC05-96OR22464 with Lockheed Martin Energy Research
Corporation, and by the University of Virginia under interagency agreement
DOE No. ERD-96-XJ003. 


\begin{thebibliography}{99}
\bibitem{notes} It can be verified that
$\left(\frac{dF_2}{dt}\right)_{\rm col}$ vanishes
to the linear order in the external electric field.
\end{thebibliography}
\end{document}